\documentclass[runningheads]{llncs}
\usepackage[T1]{fontenc}
\usepackage{graphicx}
\usepackage{multirow}
\usepackage{todonotes}
\usepackage{xcolor}
\usepackage{comment}
\usepackage{adjustbox}
\usepackage{amsmath}
\usepackage[utf8]{inputenc}
\usepackage[english]{babel}
\usepackage{color}
\usepackage{amsfonts}
\usepackage{url}
\usepackage[mathscr]{euscript}
 \let\mathscr\relax% just so we can load this and rsfs
\usepackage[scr]{rsfso}
\usepackage{caption}
\usepackage[labelfont=bf]{caption}
\usepackage{tabularx,booktabs}
\usepackage{chngcntr}
\usepackage{newfloat}
\usepackage{multirow}
\usepackage{multicol}
\usepackage{adjustbox}
\DeclareFloatingEnvironment[name={Suuplementary Figure}]{suppfigure}

\DeclareFloatingEnvironment[name={Suuplementary Table}]{suppTable}

\usepackage{multicol}
\usepackage{algorithm2e}
\usepackage{booktabs,ragged2e}
\usepackage[utf8]{inputenc}
\usepackage{lineno}
\usepackage{makecell}

\usepackage[flushleft]{threeparttable}
\usepackage{subcaption}
\usepackage{color}

\begin{document}
%-------------------
% \title{TGANet: Text-guided attention to enhance attribute-based features for generalizable polyp segmentation}
\title{TGANet: Text-guided attention for improved polyp segmentation}
\titlerunning{Text guided attention}

\author{Nikhil Kumar Tomar \inst{1}, Debesh Jha \inst{2}, Ulas Bagci \inst{2}, Sharib Ali \inst{3,4}}
\authorrunning{N. K. Tomar et al.}

\institute{NepAL Applied Mathematics and Informatics Institute for Research (NAAMII), Kathmandu, Nepal \and Department of Radiology, Feinberg School of Medicine, Northwestern University, Chicago, USA \and Institute of Biomedical Engineering, Department of Engineering Science, University of Oxford, OX3 7DQ, Oxford, UK
\and NIHR Oxford Biomedical Research Centre, Oxford, UK}
% \email{\{numan.celik, sharib.ali, jens.rittscher\}@gmail.com}}
% Nikhil Kumar Tomar {nikhil.tomar@naamii.org.np}
% Debesh Jha {debesh.jha@northwestern.edu}
% Ulas Bagci {ulas.bagci@northwestern.edu}, bagci@crcv.ucf.edu   
% Sharib Ali {sharib.ali@eng.ox.ac.uk}
% \institute{*}

\maketitle      
%-------------------
\begin{abstract}
%------------------
Colonoscopy is a gold standard procedure but is highly operator-dependent. Automated polyp segmentation, a precancerous precursor, can minimize missed rates and timely treatment of colon cancer at an early stage. Even though there are deep learning methods developed for this task, variability in polyp size can impact model training, thereby limiting it to the size attribute of the majority of samples in the training dataset that may provide sub-optimal results to differently sized polyps. In this work, we exploit \textit{size-related} and \textit{polyp number-related} features in the form of text attention during training. We introduce an auxiliary classification task to weight the text-based embedding that allows network to learn additional feature representations that can distinctly adapt to differently sized polyps and can adapt to cases with multiple polyps. Our experimental results demonstrate that these added text embeddings improve the overall performance of the model compared to state-of-the-art segmentation methods. We explore four different datasets and provide insights for size-specific improvements. Our proposed \textit{text-guided attention network} (TGANet) can generalize well to variable-sized polyps in different datasets. Codes are available at \url{https://github.com/nikhilroxtomar/TGANet}.

% through an already at the first few layers of the network that will further enable-feature weights at subsequent layers to adjust.

\end{abstract}

\section{Introduction}
Colorectal cancer (CRC) is one of the leading causes of cancer-related deaths~\cite{sung2021global} worldwide. However, high operator-dependence and subjectivity during gold standard colonoscopic procedures remain high. This is also due to the complex topology of organ, severe artefacts, constant deformation of organ, debris and stool etc. Even though cleansing of the bowel is done to improve detection rates of cancer and cancer precursor lesions such as polyps, the missed rate is still high that accounts for 26.8\% for polyps located on the right colon and 21.4\% to polyps on the left colon~\cite{kim2017miss,rex1997colonoscopic}. In addition, the missed rate for flat or sessile polyps (diminutive polyps) is grim (nearly 32.7\%). An automated system is thus clearly needed to minimize the operator subjectivity and missed rate. Semantic segmentation can classify each pixel into a class category, allowing the opportunity to learn meaningful semantic representations of polyps and their complex surroundings. Several methods do exist in the literature~\cite{fan2020pranet,jha2021real,shen2021hrenet} but most them focus on exploiting only localized spatial context. However, the nature and occurrence of polyps in colonic mucosa can be confused with colonic folds. Exploiting associated attributes such as size and occurrence (one or a few) could be used to infer and improve segmentation for hard samples.   

Encoder-decoder networks has been widely used for polyp segmentation using various modifications to boost network performance~\cite{fan2020pranet,jha2021real,shen2021hrenet}. PraNet~\cite{fan2020pranet} applied area and boundary cues in reverse attention to focus on the polyp boundary regions. The high-level feature aggregation and boundary attention blocks in the network help to calibrate some of the misaligned predictions and improve the segmentation accuracy. Similarly, HRENet~\cite{shen2021hrenet} designed an informative context enhancement (ICE) technique and adaptive feature aggregation (AFA) module and trained the model on their edge and structure consistency aware loss (ESCLoss), and obtained superior performance. Other works such as PolypSeg~\cite{zhong2020polypseg} and MSRFNet~\cite{srivastava2021msrf} uses modules incorporating multiple-scale information. An adaptive scale context module (ASCM) and semantic global context module (SGCM) was used in PolypSeg~\cite{zhong2020polypseg}. The ASCM tackles the size variations among the polyp and improves the better feature representation capability, while SGCM enhances the feature fusion between the high-level and low-level features and remove noise in the low-level features to improve the segmentation accuracy. Similarly, MSRFNet~\cite{srivastava2021msrf} integrated cross-scale fusion modules to propagate both high resolution and low-resolution features and an added shape stream network to prune polyp boundaries. 

Most of these works in the literature~\cite{fan2020pranet,shen2021hrenet,srivastava2021msrf} focuses on size variation, boundary curves, background regions, dense skip connections and dense residual scale fusions that can boost performance. However, these adjustments are made using additional layers and explicit extensions of networks and their connections. This adds to the complexity of the model that can adversely affect the generalization of test samples coming from a similar distribution and require a large dataset. In addition, it can also affect images with underrepresented polyp sizes. In this work, we propose incorporating a text guided attention mechanism using a simple byte-pair encoding for the attributes comprising polyp number and its size. In addition, we use the same encoder layer of the network to provide weights for each of these attributes. 

The main contributions of the presented work include - 1) \textit{text guided attention} to learn different features in the context of the number of polyps presence (one or many) and size (small, medium and large), 2) \textit{feature enhancement module} to strengthen the features of the encoder and pass them to the decoder, and 3) \textit{multi-scale feature aggregation} to capture features learned by different decoder blocks. We have evaluated our TGANet on four publicly available polyp datasets and compared it with five SOTA medical image segmentation methods.  

\begin{figure}[t!]
    \centering
    \includegraphics[scale=0.68]{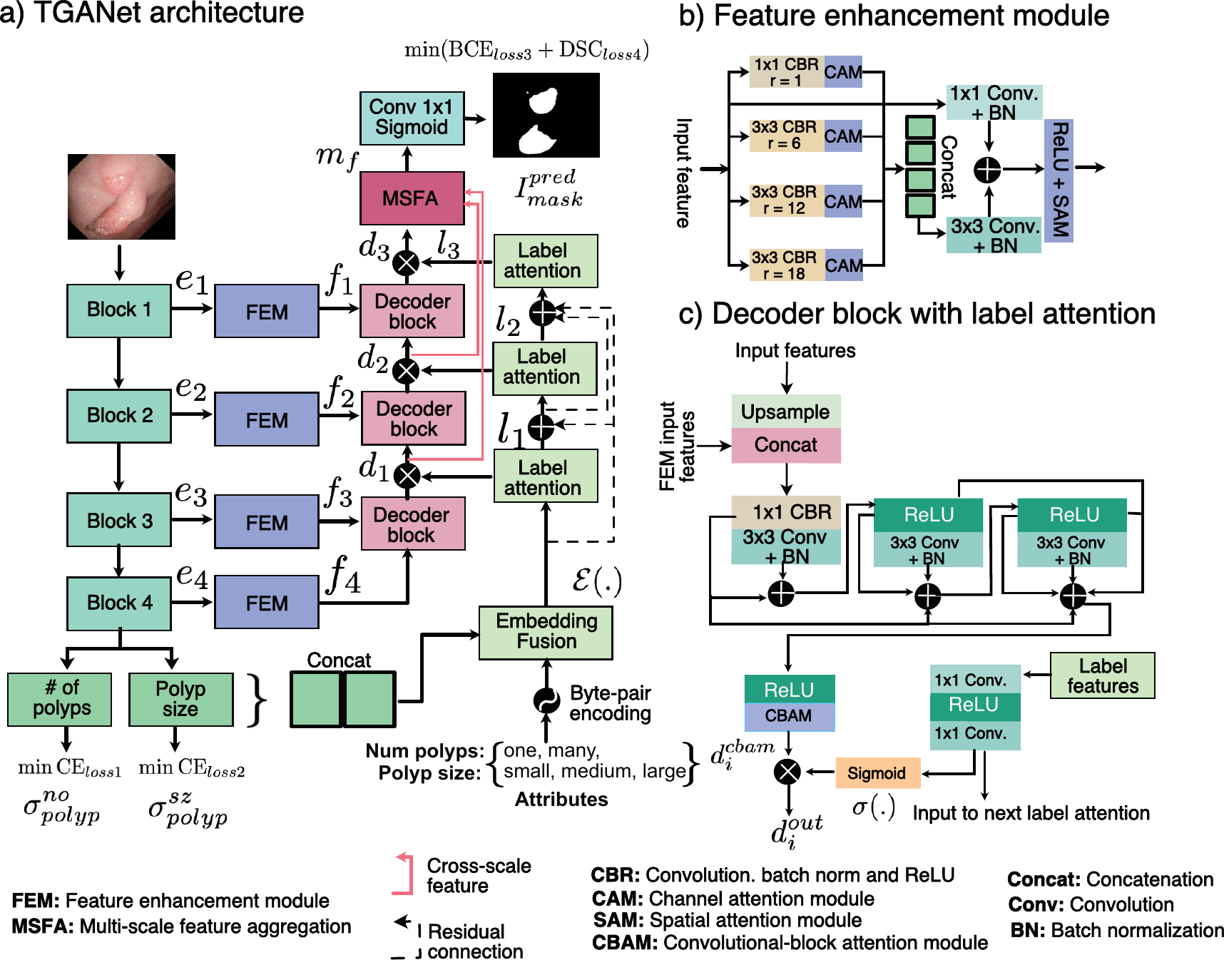}
    \caption{Block diagram of the proposed TGANet}
    \label{fig:TGANet}
\end{figure}

\section{Method}{\label{section:method}}
The proposed TGANet is a polyp segmentation architecture with text guided attention that enables to enhance feature representations such that the polyps present in images are segmented optimally independent of their size variability and occurrence. Our TGANet architecture consists of various components that are shown in Figure~\ref{fig:TGANet} and elaborated below.

\subsection{Encoder module}
TGANet is built upon a pre-trained ResNet50~\cite{he2016deep} as backbone encoder network for which we use its four different encoding blocks,  $e_{i}, i\in{1,2,3,4}$. These blocks are consecutively used for our auxiliary attribute classification task and for main polyp segmentation task. For the text-attribute classification, we use the output from the fourth encoder block as two classification task modules separately, i.e., number of polyps (one or many) and their size (small, medium and large). Here, softmax probabilities $\mathrm{\sigma}_{polyp}^{no}(.)$ and $\mathrm{\sigma}_{polyp}^{sz}(.)$ are predicted. For the main segmentation task, we take the output from each ResNet50 block and passes them through the feature enhancement module (FEM,  $f_{i}, i\in{1,2,3,4}$) that is responsible for strengthening the features by applying multiple dilated convolutions and an attention mechanism.

\subsection{Feature enhancement module}
Feature enhancement module (FEM) (see Figure~\ref{fig:TGANet} (b)) begins with four parallel dilated convolutions $\mathrm{Conv}$ with a dilation rate $r = \{1,6,12,18\}$. Each dilation is followed by a batch normalization $\mathrm{BN}$ and a rectified linear unit $\mathrm{ReLU}$ which we refer as $\mathrm{CBR}$. The output features are passed through a channel-attention module $\mathrm{CAM}$~\cite{woo2018cbam} to capture the explicit relationship between the feature channels. The highlighted features from these four dilated convolutions are then concatenated and passed through a $\mathrm{Conv}_{3\times3}$ followed by $\mathrm{BN}$ layer and added with the original input features through a $\mathrm{Conv}_{1\times1}$. The resulting features are then followed by a $\mathrm{ReLU}$ activation function, and a spatial attention mechanism $\mathrm{SAM}$~\cite{woo2018cbam} is applied to suppress the irrelevant regions.

\subsection{Label attention}
Label attention module is designed to provide learned text-based attention to the output features of the decoder blocks in our TGANet. Here, we use three label attention modules, $l_{i}, i\in{1,2,3}$, as soft channel-wise attention to the three decoder outputs that enables larger weights to the representative features and suppress the redundant ones. The first label attention module uses the output of the embedding fusion $\mathcal{E}(.)$ obtained by element-wise dot product between the softmax probability concatenation $\{\mathrm{\sigma}_{one}, \mathrm{\sigma}_{many}, \mathrm{\sigma}_{small}, \mathrm{\sigma}_{medium}, \mathrm{\sigma}_{large} \}$ with the encoded text embedding. Say, $\mathcal{A}=\{{\mathrm{one}, \mathrm{many}, \mathrm{small}, \mathrm{medium}, \mathrm{large}}\}$ be the attributes that are encoded using byte-pair encoding~\cite{heinzerling2018bpemb} and denoted by $\mathcal{A}_{encode}$ with $\{a_{j}^{k}\}$ as vector embedding for each attribute $j$ of length $|k|$, then $\mathcal{E}(.)$ that is given by:
\begin{equation}
    \mathcal{E} = \sigma_j \odot {a_{j}^{k}}, \quad  \forall{k}.
\end{equation}

The output of the label attention module is referred to as label features $l_{f}$ in this paper. 

\subsection{Decoder}
The decoder in the proposed TGANet is comprised of three different decoder blocks $d_{i}, i\in{1,2,3}$, of which each takes the input features to upsample it and pass it through some convolutional layers to produce the output. This output is refined using the label attention module $l_i$ and passed to the subsequent decoder blocks $d_{i}$ (see Figure~\ref{fig:TGANet} (c)). The first decoder block takes the output of the fourth FEM  $f_{4}$ to upsample it using bilinear interpolation by a factor of two, and then it is concatenated with the output features from the third FEM $f_{3}$. The resulting concatenated feature is passed through a $\mathrm{Conv}_{1\times1}$-$\mathrm{BN}$-$\mathrm{ReLU}$ referred as $\mathrm{CBR}$ followed by a sequence of three $\mathrm{Conv}_{3\times3}$-$\mathrm{BN}$, further accompanied by their multiple residual connections and a $\mathrm{ReLU}$ activation function with subsequent convolutional block attention module represented as $d_i^{cbam}$. An element-wise multiplication is done to allow additional soft-attention from the computed label features $l_{f}$ using a sigmoid function for each decoder block output $d_i^{out}, i \in {1,2,3}$ given by:
\begin{equation}
    d_i^{out} = d_i^{cbam} \odot \sigma([\mathrm{Conv}-\mathrm{ReLU}-\mathrm{Conv}]l_{f}), \quad \forall{i\in {1,2,3}}
\end{equation}

\subsection{Multi-scale feature aggregation}
Multi-scale feature aggregation (MSFA) module (see supplementary Figure 1) is used to fuse multi-scale feature representations at various decoder outputs $d_i^{out}, i\in{1,2,3}$ that allows to capture learned features. We take the first two features $\{d_1^{out}, d_2^{out}\}$ and pass them through a bilinear upsampling to ensure that all three features have the exact spatial dimensions followed by linear $1\times 1$ convolution layers, $\mathrm{BN}$ and $\mathrm{ReLU}$ activation before concatenation. To boost the capture of non-linear features we further apply a series of convolutional layers, $\mathrm{BN}$ and $\mathrm{ReLU}$ together with multiple residual connections for improved flow of information. The output is represented as $m_f$ which is responsible for our predicted segmentation map $I^{pred}_{mask}$ given by: $I^{pred}_{mask} = \sigma(\mathrm{Conv}_{1\times 1}(m_f))$.

\indent{\textbf{Joint loss optimization:}} We jointly minimize loss for both the auxiliary classification tasks (cross-entropy losses, $\mathrm{CE}_{loss1}$, $\mathrm{CE}_{loss2}$) and the segmentation task (binary cross entropy, $\mathrm{BCE}_{loss3}$ and dice loss, $\mathrm{DSC}_{loss4}$) with equal weights.
\section{Experiments and results}
\begin{table}[t!]
\footnotesize
\centering
\caption{Quantitative results on the experimented polyp datasets.}
 \begin{tabular} {l|c|c|c|c|c|c}
\toprule
\textbf{Method}  &\textbf{Backbone} & \textbf{mIoU}  &\textbf{mDSC}  &\textbf{Recall}& \textbf{Precision} &\textbf{F2} \\ 
\hline

\multicolumn{7}{@{}l}{\textbf{Dataset: Kvasir-SEG~\cite{jha2020kvasir}}}                   \\ \hline
U-Net~\cite{ronneberger2015u} &	-&	0.7472&	0.8264&	0.8504&	0.8703&	0.8353 \\
%U-Net++~\cite{zhou2018unet++}&	-&	0.742&	0.8228&	0.8437&	0.8607&	0.8295 \\
%ResU-Net++~\cite{jha2019resunet++}&	-&	0.5341&	0.6453&	0.6964&	0.7080&	0.6576 \\
HarDNet-MSEG~\cite{huang2021hardnet}&	HardNet68&	0.7459&	0.8260&	0.8485&	0.8652&	0.8358 \\
ColonSegNet~\cite{jha2021real}&	-&	0.6980&	0.7920&	0.8193&	0.8432&	0.7999 \\
DeepLabV3+~\cite{chen2018encoder}&	ResNet50&	0.8172&	0.8837&	0.9014&	0.9028&	0.8904 \\
PraNet~\cite{fan2020pranet}&	Res2Net&	0.8296&	0.8942&	0.9060&	\textbf{0.9126}&	0.8976 \\
\textbf{TGANet (Ours)}&	ResNet50&	\textbf{0.8330}&	\textbf{0.8982}&	\textbf{0.9132}&	0.9123&	\textbf{0.9029} \\
\hline

\multicolumn{7}{@{}l}{\textbf{Dataset: CVC-ClinicDB~\cite{bernal2015wm}}}                   \\ \hline
U-Net~\cite{ronneberger2015u}&	-&	0.8428&	0.8978&	0.9001&	0.9209&	0.8981 \\
%U-Net++~\cite{zhou2018unet++}&	-&	0.8337&	0.8913&	0.9129&	0.8988&	0.9026 \\
%ResU-Net++~\cite{jha2019resunet++}&	-&	0.7229&	0.8113&	0.8115&	0.8657&	0.8083 \\
HarDNet-MSEG~\cite{huang2021hardnet}&	HardNet68&	0.8388&	0.8967&	0.8929&	0.9216&	0.8938 \\
ColonSegNet~\cite{jha2021real}&	-&	0.8248&	0.8862&	0.8828&	0.9017&	0.8826 \\
DeepLabV3+~\cite{huang2021hardnet}&	ResNet50&	0.8973&	0.9391&	\textbf{0.9441}&	0.9442&	0.9389 \\
PraNet~\cite{fan2020pranet}&	Res2Net&	0.8866&	0.9318&	0.9347&	0.9479&	0.9333 \\
\textbf{TGANet (Ours)}& ResNet50&		\textbf{0.8990}&	\textbf{0.9457}&	0.9437&	\textbf{0.9519}&	\textbf{0.9439} \\
\hline

\multicolumn{7}{@{}l}{\textbf{Dataset: BKAI~\cite{lan2021neounet}} \label{tab:bkai}}    \\
\hline            
U-Net~\cite{ronneberger2015u}&	-&	0.7599&	0.8286&	0.8295&	0.8999&	0.8264 \\
%U-Net++~\cite{zhou2018unet++}&	-&	0.7563&	0.8275&	0.8388&	0.8942&	0.8308 \\
%ResU-Net++~\cite{jha2019resunet++}&	-&	0.6280&	0.7130&	0.72490&	0.8578&	0.7132 \\
HarDNet-MSEG~&	HardNet68&	0.6734&	0.7627&	0.7532&	0.8344&	0.7528 \\
ColonSegNet~\cite{jha2021real}&	-&	0.6881&	0.7748&	0.7852&	0.8711&	0.7746 \\
DeepLabV3+~\cite{chen2018encoder}&	ResNet50&	0.8314&	0.8937&	0.8870&	\textbf{0.9333}&	0.8882 \\
PraNet~\cite{fan2020pranet}&	Res2Net&	0.8264&	0.8904&	0.8901&	0.9247&	0.8885 \\
\textbf{TGANet (Ours)}&	ResNet50&\textbf{0.8409}&	\textbf{0.9023}&	\textbf{0.9026}&	0.9208&	\textbf{0.9002} \\
\hline

\multicolumn{7}{@{}l}{\textbf{Dataset: Kvasir-Sessile~\cite{jha2021comprehensive}}  \label{tab:sessile}}            \\  \hline
U-Net~\cite{ronneberger2015u}&	-&	0.2472&	0.3688&	0.7237&	0.3264&	0.4635 \\
%U-Net++~\cite{zhou2018unet++}&	-&	0.2448&	0.3761&	0.5361&	0.4086&	0.4225 \\
%ResU-Net++~\cite{jha2019resunet++}&	-&	0.1763&	0.2754&	0.9729&	0.1781&	0.4349 \\
HarDNet-MSEG~&	HardNet68&	0.1565&	0.2558&	0.5403&	0.2236&	0.3298 \\
ColonSegNet~\cite{jha2021real}&	-&	0.2113&	0.3278&	0.5234&	0.3336&	0.3868 \\
DeepLabV3+~\cite{chen2018encoder}&	ResNet50&	0.5927&	0.7078&	0.7085&	0.8225&	0.7009 \\
PraNet~\cite{fan2020pranet}&	Res2Net&	0.6671&	0.7736&	0.8069&	0.8244&	0.7871 \\
\textbf{TGANet (Ours)}&	ResNet50& \textbf{0.6910}& \textbf{0.7980}& \textbf{0.7925}& \textbf{0.8588}& \textbf{0.7879} \\

\hline\multicolumn{7}{@{}l}{\textbf{Training dataset: Kvasir-SEG -- Test dataset: CVC-ClinicDB} \label{tab:cross-data}}            \\  \hline
U-Net~\cite{ronneberger2015u}&	-&	0.5433&	0.6336&	0.6982&	0.7891&	0.6563 \\
%U-Net++~\cite{zhou2018unet++}&	-&	0.5475&	0.6350&	0.6933&	0.7967&	0.6556 \\
%ResU-Net++~\cite{jha2019resunet++}&	-&	0.3585&	0.4642&	0.5880&	0.5770&	0.5084 \\
HarDNet-MSEG~\cite{huang2021hardnet}&	HardNet68&	0.6058&	0.6960&	0.7173&	0.8528&	0.7010 \\
ColonSegNet~\cite{jha2021real}&	-&	0.5090&	0.6126&	0.6564&	0.7521&	0.6246 \\
DeepLabV3+~\cite{chen2018encoder}&	ResNet50&	0.7388&	0.8142&	\textbf{0.8331}&	0.8735&	0.8198 \\
PraNet~\cite{fan2020pranet}&	Res2Net&	0.7286&	0.8046&	0.8188&	0.8968&	0.8077 \\
\textbf{TGANet (Ours)}&	ResNet50&	\textbf{0.7444}& \textbf{0.8196}& 0.8290&	\textbf{0.8879}& \textbf{0.8207} \\
\bottomrule
\end{tabular}
\label{tab:results}
\vspace{-5mm}
\end{table}

\subsection{Datasets}
To evaluate the performance of our TGANet, we have used four publicly available polyp segmentation benchmark datasets including Kvasir-SEG~\cite{jha2020kvasir}, CVC-ClinicDB~\cite{bernal2015wm}, BKAI\cite{lan2021neounet}, and Kvasir-Sessile~\cite{jha2021comprehensive} (details are presented in supplementary Table 1). Relevant to our experiment, Kvasir-Sessile~\cite{jha2021comprehensive} contains 196 small, diminutive, sessile and flat polyps that are less than 10 mm in size. 

\subsection{Implementation Details}
All models are trained on NVIDIA GeForce RTX 3090 GPU, and images are resized to $256 \times 256$ pixels with 80:10:10 training, validation, and testing splits except for Kvasir-SEG, where we adopted the official split of 880/120 for training and testing. Simple data augmentation strategy including random rotation, vertical flipping, horizontal flipping, and coarse dropout are used. All models are trained on a similar hyperparameters configuration with a learning rate of $1e^{-4}$, batch size of 16, and optimized with Adam optimizer. An early stopping mechanism and ReduceLROnPlateau is used to prevent models from overfitting.  

Standard medical image segmentation metrics such as mean intersection over union (mIoU), mean Sørensen–dice coefficient (mDSC), recall, precision, F2-score and frame per second (FPS) are used. 

\subsection{Results}
\vspace{-2mm}
\begin{table}[t!]
    \centering
\caption{mDSC for different \textit{sizes} and \textit{polyp counts on Kvasir-SEG}}~\cite{jha2020kvasir}
    \begin{tabular}{ l| l| l| l| l| l} 
\toprule
Method &small & medium & large &one& many\\ \midrule
DeepLabV3+~\cite{chen2018encoder}  &0.8776 &0.9003  &0.8633  &0.8922 &0.8289  \\
PraNet~\cite{fan2020pranet} &0.8826  &0.9079 &\textbf{0.8900}  &0.9071  &0.8106   \\
TGANet  &\textbf{0.8869}  &\textbf{0.9203} &0.8769  &\textbf{0.9075}  &\textbf{0.8378}   \\
\bottomrule
    \end{tabular}
    \label{tganet-effect}
    \end{table}
\begin{figure} [!t]
    \centering
    \includegraphics[width=0.7\textwidth ]{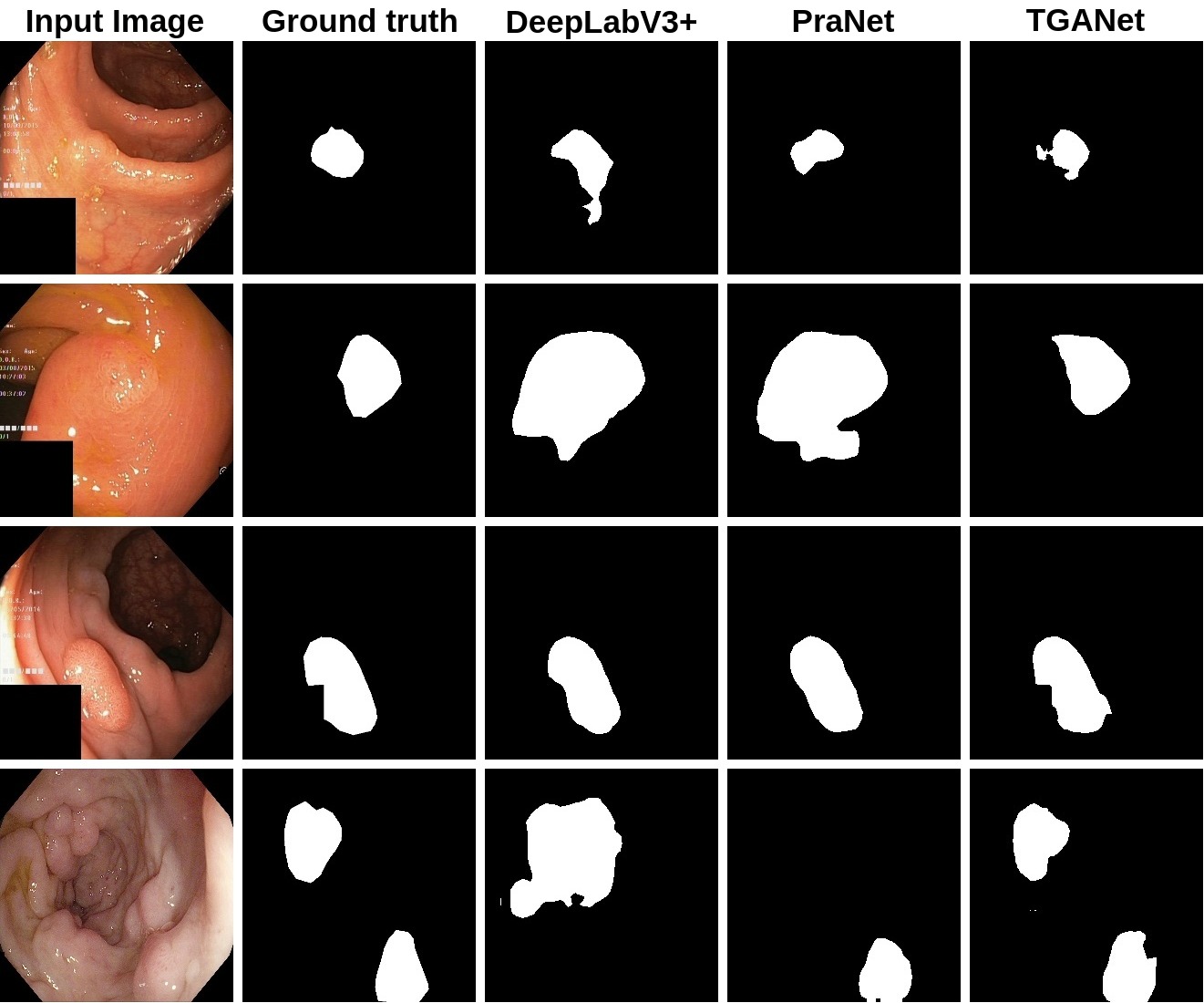} 
    \caption{Qualitative results comparison on the Kvasir-SEG dataset.}
    \label{fig:qualitative}
    \vspace{-5mm}
\end{figure}
We have compared our results with five SOTA methods that include UNet~\cite{ronneberger2015u}, HarDNet-MSEG~\cite{huang2021hardnet}, ColonSegNet~\cite{jha2021real}, DeepLabv3+~\cite{chen2018encoder}, and PraNet~\cite{fan2020pranet}. These algorithms are widely used baselines in both polyp segmentation and general medical image segmentation. The quantitative results are presented in Table~\ref{tab:results}.
\vspace{-3mm}
% For medical image segmentation tasks,  mIoU and mDSC are widely accepted metrics for comparing and evaluating machine learning models.
\paragraph{\textbf{Results on Kvasir-SEG:}}
Table~\ref{tab:results} shows that TGANet outperforms all the SOTA methods with a mIoU of 0.8330 and mDSC of 0.8982. Our TGANet outperforms most competitive PraNet~\cite{fan2020pranet} by 1.58\% in mIoU and 1.45\% mDSC. 
\vspace{-3mm}
\paragraph{\textbf{Results on CVC-ClinicDB:}}
For CVC-ClinicDB dataset, TGANet outperforms all SOTA methods reporting the highest mIoU and mDSC of 0.8990 and 0.9457, respectively. Our method outperformed the most competitive DeepLabV3+~\cite{chen2018encoder} with a mIoU of 0.17\% and mDSC of 0.66\%.   
\vspace{-3mm}
\paragraph{\textbf{Results on BKAI:}}
Table~\ref{tab:results} shows the comparison of the result on the BKAI dataset that show that our proposed TGANet obtains mIoU of 0.8409 and mDSC of 0.9023 and outperforms the best performing DeepLabV3+~\cite{chen2018encoder} by 0.95\% on mIoU and 0.86\% on mDSC. 
\vspace{-3mm}
\paragraph{\textbf{Results on Kvasir-Sessile:}}
Kvasir-Sessile dataset is clinically most relevant as it has flat and sessile polyps. On this dataset, it can be observed (see Table~\ref{tab:results}) that our TGANet surpasses all the other methods in all the evaluation metrics. It outperforms the best performing PraNet~\cite{fan2020pranet} by a large margin of 2.39\% on mIoU and 2.44\% on mDSC. Similarly, almost 10\% improvement is observed compared to the DeepLabV3+~\cite{chen2018encoder} in this case which is a significant improvement. 
\vspace{-3mm}
\paragraph{\textbf{Results on cross dataset:}}
To explore the generalization capability of our proposed TGSNet, we train the model on Kvasir-SEG and test it on the CVC-ClinicDB. The cross-dataset test (Table~\ref{tab:results}) also suggested improvements compared to all SOTA methods and obtained an increment of 0.56\% on mIoU and 0.54\% on mDSC compared to the SOTA DeepLabv3+~\cite{chen2018encoder}. 
\vspace{-3mm}

\paragraph{\textbf{Results on size and number-based sampled polyps:}}
To show the effectiveness of our proposed TGANet, we evaluated test samples of Kvasir-SEG-based on the attributes used in training. It can be observed in Table~\ref{tganet-effect} that our model outperforms the best SOTA methods for almost all cases. For the `small', `medium' and `many cases', the improvement ranges from nearly 1-2\%. 

Our qualitative results (see Figure~\ref{fig:qualitative}) demonstrate a clear improvement of our text-based attention method for different sizes and number polyp samples. It is evident that both PraNet~\cite{fan2020pranet} and DeepLabV3+~\cite{chen2018encoder} failed to capture sample with two polyps (4th row) and also provided over segmentation for the small (1st row) and medium polyps (2nd row). Additionally, we have provided the total number of parameters, flops and FPS in {supplementary material Table 2}. 
\subsection{Ablation study}
\begin{table*}[!t]
\footnotesize
\centering
\caption{Ablation study of TGANet on Kvasir-SEG}
 \begin{tabular} {l|l|c|c|c|c|c}
\toprule
\textbf{No}  &\textbf{Method} & \textbf{mIoU}  &\textbf{mDSC}  &\textbf{Recall}& \textbf{Precision} &\textbf{F2} \\
\midrule
%
%\& TGANet w/o label&	0.8270&	0.8916&	0.9085&	0.9093&	0.8976 \\
\#1 & TGANet w/o label and classifier&	0.8104&	0.8786&	0.8987&	0.8970&	0.8850 \\
\#2 & TGANet w/o MSFA & 0.8151&	0.8832&	0.9061&	0.8999&	0.8907 \\

\#3& TGANet w/o FEM &0.8084& 0.8766&	0.8968&	0.9010&	0.8838 \\

\#4& \shortstack{TGANet w/o (label+classifier+\\MSFA+FEM)}&		0.8063&	0.8747&	0.8963&	0.8971&	0.8798	\\

\#5& \textbf{TGANet (Ours)}&	\textbf{0.8330} &\textbf{0.8982}&	\textbf{0.9132}&	\textbf{0.9123}&	\textbf{0.9029} \\ 
\bottomrule
\end{tabular}
\label{ablation}
\end{table*}

To validate the effectiveness and importance of the core components used in the network, we compare TGANet with its five variants, which is presented in Table~\ref{ablation}. The results suggest that the introduction of the text guided attention along with the label boosts the performance of the network. The results show that TGANet improves the baseline without the label and classifier (\#1) by 2.26\% on mIoU and 1.96\% on mDSC.   

\section{Conclusion}
We proposed a text-guided attention architecture (TGANet) to tackle polyps' variable size and number for robust polyp segmentation. We have used multiple feature enhancement modules connected with different encoder blocks to achieve this. An auxiliary task is learned together with the main task to compliment both the size-based and number-based feature representations and used as label attentions in the decoder blocks. Additionally, the multi-scale fusion of the features at the decoder enabled our network to deal with these attribute changes. Our experimental results demonstrated the effectiveness of our TGANet outperformed and provided higher segmentation performance on flat and sessile polyps that are clinically important. 
% \subsubsection*{Acknowledgement}
%
\bibliographystyle{splncs04}
\bibliography{ref}

\section*{Supplementary material}
\begin{suppfigure}[h!]
    \centering
    \includegraphics[scale=0.4]{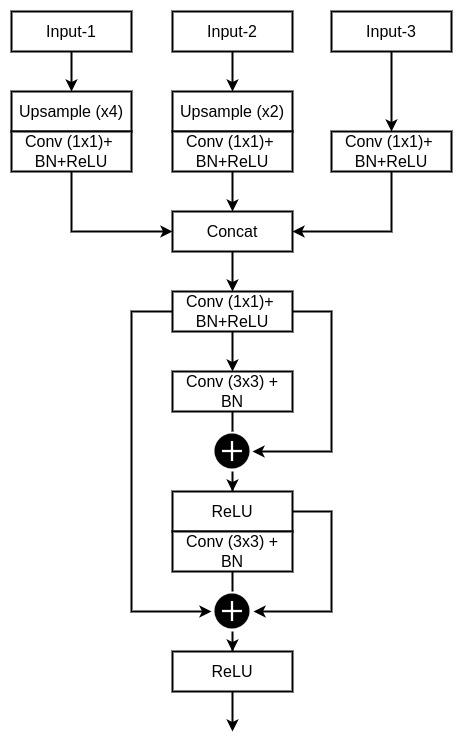}
    \caption{Block diagram of the proposed multi-scale fusion aggregation module that takes different scale features from three decoder blocks and concatenates them using upsampling, convolution, batch normalization and $\mathrm{ReLU}$ activations. Also, we apply residual-connections for different layers as shown.}
    \label{fig:TGANet}
\end{suppfigure}

\begin{suppTable} [h!]
 \caption{Polyp datasets used in our experiments with number of images, input size and their availability. For some indicated dataset test data are not available (e.g. BKAI) for which we used training data and split it into 80:10:10 for train, validation and test. Also, we use small, flat and sessile polyps in Kvasir-SEG.}
    \label{table:datasettable}
    \centering
          \begin{tabular}{ l| c| c| c} 
                \toprule
                Dataset & \shortstack{Images} & Input size & Availability\\ 
              \bottomrule
Kvasir-SEG  &1000 &Variable &Public\\
CVC-ClinicDB  &612 &384 × 288 &Public\\
BKAI &1000$^\dag$ &Variable &Public\\ 
Kvasir-Sessile$^\diamond$ &196 &Variable &Public\\ 
% ETIS Larib Polyp DB~\cite{silva2014toward}&196 & $1225 \times 966$ &Public \\ 
\bottomrule
\multicolumn{4}{l}{$^\dag$ test data unavailable} \hspace{.1cm} $^\diamond${sessile polyps from Kvasir-SEG}
\end{tabular}
\end{suppTable}

\begin{suppTable}[h!]
    \centering
     \caption{Algorithm complexity of methods used in the study. Number of model parameters, flops, image size and frame-per-second is provided for both state-of-the-art (SOTA) methods and proposed TGANet.}
    \begin{tabular}{l|l|l|l|l} 
\toprule
Method  &Parameters (Million) &Flops (GMac) & Image Size & FPS \\ \midrule 
U-Net&31.04 &54.75 &$256\times 256$ & 164.93\\
%U-Net++&9.16 &34.65 &$256\times 256$ & 124.72\\
%ResU-Net++&4.06 &15.81 &$256\times 256$ &53.10\\
HarDNet-MSEG&33.34 &6.02 &$256\times 256$ &42.95\\
ColonSegNet&5.01 &62.16 &$256\times 256$ &127.95 \\ 
DeepLabV3+&39.76	&43.31 & $256\times 256$&99.79\\
PraNet&32.55	&6.93 & $256\times 256$& 36.90\\
\textbf{TGANet (Ours)} &19.84 &41.88 & $256\times 256$ & 37.64\\
\bottomrule
    \end{tabular}
    \label{algorithm_complexity}
    \end{suppTable}
\end{document}